# Studies of Electron Avalanche Behavior in Liquid Argon

J.G. Kim, S.M. Dardin, K.H. Jackson, R.W. Kadel, J.A. Kadyk, V. Peskov, W.A. Wenzel

*Abstract*-- **Electron avalanching in liquid argon is being studied as a function of voltage, pressure, radiation intensity, and the concentrations of certain additives, especially xenon. The avalanches produced in an intense electric field at the tip of a tungsten needle are initiated by ionization from a moveable americium ($^{241}$Am) gamma ray source. Photons from xenon excimers are detected as photomultiplier signals in coincidence with the current pulse from the needle. In pure liquid argon the avalanche behavior is erratic, but the addition of even a small amount of xenon (£100ppm) stabilizes the performance. Similar attempts with neon (30%) as an additive to argon have been unsuccessful. Tests with higher energy gamma rays ($^{57}$Co) yield spectra and other performance characteristics quite similar to those using the $^{241}$Am source. Two types of signal pulses are commonly observed: a set of pulses that are sensitive to ambient pressure, and a set of somewhat smaller pulses that are not pressure dependent.**

## I. INTRODUCTION

The potential advantages of liquids over gases for particle detection have stimulated research on the noble liquids [1]. The intrinsic spatial resolution in liquids is better, both because the density of ions from charged particles is much greater, and because the diffusion of drifting electrons is less. Moreover, for experiments with small particle cross-sections (notably neutrino experiments), the density of the target detection medium is of utmost importance. In previous detectors using drifting electrons in liquid argon (LAr) [2], signals produced without avalanching are small with a low (10:1) signal/noise ratio. The spatial resolution was at the 150μm level. However, if significant avalanche gain (≥100) can be produced with this type of LAr detector, much better resolution should be possible. Reliable avalanche behavior was achieved with high purity xenon [3], with a proportional response much as in gas avalanche counters. But the results with more affordable pure argon were quite different: avalanche production was erratic, unreliable, and the detectors were unstable and pressure dependent.

The present investigation was directed toward finding a more satisfactory mode of operation for a liquid argon avalanche detector. One motivation is the potential application to a very large scale (≥ 10kton) neutrino factory experiment to observe the τ lepton, which is short-lived and would leave a very short track. Very good spatial resolution is essential, requiring a point resolution of ~10μm for every ~100μm of track length. The avalanche detector would then serve as both as a massive neutrino target and as a high resolution detector at an acceptable cost.

In the present work, avalanches have been observed in liquid argon/xenon mixtures, and studied as a function of voltage, pressure, radiation intensity, and gamma-ray energy. A very sharp needle (~0.25μm tip radius) has been used as the anode to achieve the field enhancement on the needle tip required for electron avalanching. The concentration of xenon was varied to investigate the stability of operation and pulse amplitude and even small Xe/Ar concentrations were found to stabilize the performance.

## II. EXPERIMENTAL SETUP

The "test vessel" is a stainless steel cylinder of about 6 liter total volume (Fig. 1). It has several ports for: (a) a moveable 0.1Ci $^{241}$Am source that illuminates the needle through a 1mm thick stainless steel window; (b) a vacuum UV PMT [4] to detect photons associated with the avalanche through a CaF$_2$ lens and MgF$_2$ window; (c) a gaseous (P10 or CF$_4$) detector with a CsI photocathode also for UV photon detection; and (d) a thin aluminum window (0.25mm thick) through which an alternative $^{57}$Co source can illuminate the needle. A level indicator for LAr, based upon a custom capacitance probe, measures the LAr level to a few millimeters. The temperature inside is monitored by a low-temperature thermistor and by four platinum resistors mounted ~1cm apart vertically. The

Manuscript received Nov 3, 2001. This work was supported by the Director, Office of the Energy Research of the U.S. Department of Energy under Contract No. DE-AC03-76SF00098.

J. G. Kim is with the Physics Division, Lawrence Berkeley National Laboratory, Berkeley, CA 94720 USA on leave from Physics Department, MyongJi University, Young-In, Korea 449-728 (telephone: 510-486-4579, e-mail: JGKim@lbl.gov).

S. M. Dardin is with the Physics Division, Lawrence Berkeley National Laboratory, Berkeley, CA 94720 USA (telephone: 510-486-6598, e-mail: SMDardin@lbl.gov).

K. H. Jackson is with the Material Science Division, Lawrence Berkeley National Laboratory, Berkeley, CA 94720 USA (telephone: 510-486-6894, e-mail: KHJackson@lbl.gov).

R. W. Kadel is with the Physics Division, Lawrence Berkeley National Laboratory, Berkeley, CA 94720 USA (telephone: 510-486-7360, e-mail: RWKadel@lbl.gov).

J. A. Kadyk is with the Physics Division, Lawrence Berkeley National Laboratory, Berkeley, CA 94720 USA (telephone: 510-486-7189, e-mail: JAKadyk@lbl.gov).

V. Peskov is with the Royal Institute of Technology, Stockholm, Sweden (telephone: 48-8-5537-8182, e-mail: Vladimir.Peskov@cern.ch ).

W. A. Wenzel is with the Physics Division, Lawrence Berkeley National Laboratory, Berkeley, CA 94720 USA (telephone: 510-486-6918, e-mail: B_Wenzel@lbl.gov).

temperature of the liquid in the test vessel is typically ~87°K, while the UV PMT and CsI photodetectors operate at room temperature and pressure.

The test vessel is cooled using an open bath of liquid argon, with some liquid nitrogen pre-cooling. To minimize impurities and outgassing, the test vessel and associated gas filling system are periodically vacuum baked at ~120°C for more than 24 hours. The base level vacuum is about $1 \times 10^{-5}$ Torr. To achieve the purity level needed to drift electrons, the boil-off gas from the LAr supply dewar (99.99% pure) is passed through three filters: a Matheson Molecular Sieve 4A [5], an "Oxysorb" $O_2$ filter [6], and a Aeronex "Gatekeeper" inert gas purifier including a 0.5μm particulate filter [7]. After filtering, the Ar gas has a measured oxygen impurity level less than about 10 ppb based on measurements of our $O_2$ meter [8]; it is expected to be about 1ppb, based upon the published filter specifications.

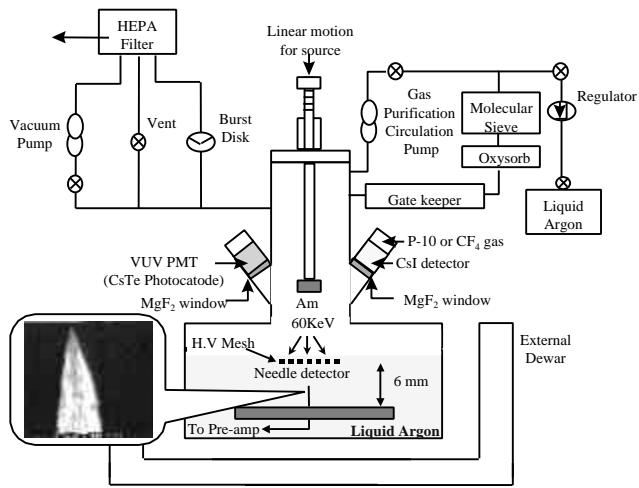

Fig. 1. A schematic of the test vessel in which avalanche tests were done using liquid argon, and an photograph of a needle tip using an electron microscope. Not shown is the $O_2$ meter in the input line, between the test vessel and the Gatekeeper gas purifier, or the port for the $^{57}$Co source.

Condensation of ~2.6 liters of LAr requires 1~2 hours after the test vessel is cold. When the level indicator showed that liqufied argon has reached some intermediate level a measured quantity of the high purity gas mixture (Ar 95%/ Xe 5%) from a gas cylinder is introduced directly into the test vessel to obtain the desired Xe/Ar concentration. Chemical analysis of the input Xe/Ar gas mixture shows the $O_2$ concentration of the gas to be 2.6ppm, and an additional search for electronegative impurities using and ECD (Electron Capture Detector) detected no impurities [9]. Any contamination of the Xe/Ar gas is further diluted by a factor of ~100 when it is added to the LAr in the test vessel.

After adding the Xe/Ar mixture, additional Ar gas is condensed to complete the 2.6 liter fill. After the filling is completed, the Ar gas supply is shut off, and the test vessel is pressurized to about 1/3 atm (5psig) of He gas to suppress boiling in the test vessel. The He pressure can be increased to allow testing pressure up to 1.3 atm (20psig). The He gas is filtered through the same filters as the Ar gas.

Without the He gas the operating point of the (closed) system is approximately atmospheric pressure. Pressurizing the system with Ar (at constant temperature) does not allow the system to be operated at higher pressure because the extra argon gas quickly liquifies. Hence we add He to maintain a slight overpressure, or to increase the pressure up to ~1.3 atm.

For some of the tests, an alnico bar magnet mounted on a pivot near the bottom of the test vessel is rotated under the influence of an external driving magnet. This system mixed the liquid to insure a uniform Xe/Ar concentration. The quoted Xe concentrations assume that all the Xe dissolves in the LAr.

As can be seen in Fig. 1, an electric field is produced in a gap of 6 mm between the wire mesh cathode and a ground plane; both electrodes are made of stainless steel. The mesh is operated typically at a potential of 1 to 3 kV. Protruding upward through a hole in the ground plane is an electrically isolated tungsten needle, also at dc ground potential. The needle tip is approximately halfway between the mesh and ground plane. The needle tapers from a diameter of 0.25mm over most of its length, to ~0.5μm diameter near the rounded tip. The sensitive region between the needle and mesh is illuminated by 60keV gammas from the 0.1Ci $^{241}$Am source (α's do not penetrate the 1mm thick SS window), which is moveable between 5cm and 15cm above the 50% optically transparent HV mesh.

Direct currents from the mesh, needle, ground plane, and source holder are monitored and recorded using picoammeters [10] via the LabView program [11]. Pulses from the needle are observed on an oscilloscope, and both avalanche rates and pulse heights are monitored and stored using an Oxford Instruments pulse height analyzer (PHA) [12]. The signal from the needle is amplified first with a charge sensitive preamplifier [13], and then with an LBNL shaper/amplifier ("TranLamp"), adjusted to have 5μsec differentiation and integration time constants for pulse height measurements. The total gain (preamp and TranLamp) is about 11V/pC. The time spectra are obtained by adjusting the integration/differentiation time constants to 0.1μs/0.2μs, respectively, and using the resultant current pulses from needle avalanches as the start signal. The PMT pulses provided the stop signals.

## III. RESULTS

### A. Needle Current

Electron-ion pairs are created by ionizing radiation from the $^{241}$Am source in the liquid argon. The positive ions with very low mobility (~$6 \times 10^{-4}$ cm²/V-s, [14]) are collected on the cathode mesh. The electrons with a much larger velocity, ~3 to 5 mm/μs, are drawn to either the needle or the ground plane. The electric field at the narrow needle tip is a factor of nearly three thousand times the ambient field in the gap.

Below the avalanche threshold, small signals are observed on the scope just above the noise level. These are electrons collected from the primary ionization events, i.e., without avalanche multiplication. Above a threshold field, electron avalanching occurs. This is observable both as an increase in needle current and by a large increase in pulse height as seen on the scope and measured on the PHA.

## B.  Prolate Spheroid Needle Model

A close approximation to the actual needle geometry is that of a prolate spheroid of large aspect ratio, $p = \sqrt{(A/T)}$, where A =3mm is the height of the needle and T=0.25μm, is the radius of the tip (Fig. 2). Smythe [15] gives the potential V(Z,ρ) for the upper half prolate spheroid extending above a ground plane in a uniform electric field $E_0$; Z and ρ are the cylindrical coordinates with Z taken along the major axis of the prolate spheroid. The potential V is given by the expression:

$$V = E_o Z[1-(\coth^{-1}\boldsymbol{h}-\frac{1}{\boldsymbol{h}})/(\coth^{-1}\boldsymbol{h}_o-\frac{1}{\boldsymbol{h}_o})], \text{ where}$$

$$Z = C\boldsymbol{h}\boldsymbol{z} , \ \boldsymbol{r} = C\sqrt{(\boldsymbol{h}^2-1)(1-\boldsymbol{z}^2)}, \ A^2-B^2=C^2, \quad (1)$$
$$\text{and } \boldsymbol{h}_O = A/C.$$

In this notation, the needle is a surface of constant η=η₀, and the tip of the needle is at ζ=ζ₀=1. The electric field (E(z), z = Z/C) along the Z axis is:

$$E(z)/E_o = 1 + [\frac{z}{z^2-1} - \coth^{-1}z]/[\coth^{-1}\boldsymbol{h}_o-\frac{1}{\boldsymbol{h}_o}]. \quad (2)$$

For the large aspect ratio of the needle considered here, these expressions can be simplified to yield the electric field at the tip of the needle as: $E(\boldsymbol{h}_o,\boldsymbol{z}_o)=VA/(kTH)$, where k equals $\ln(2p)$-1≈4.4 for the geometry of Fig. 2. The electric field on the mesh directly above the needle is given by $E(\text{mesh})=E_o(1+2\times0.03)=1.06 E_o$, where the factor 2 comes from the image charge of the needle in the mesh, and 0.03 is the value at the mesh of the terms in square brackets in (2). The result is that the field at the tip is enhanced by a factor of 2670 over the ambient field between the mesh and the ground plane for the dimensions of Fig. 2. Based upon the known mesh voltage, and our best knowledge of the gap size and needle position and tip radius, we have calculated the electric field at the needle tip corresponding to the measured avalanche threshold, both analytically as described above and using a simulation program [16]. These methods give reasonable agreement for an estimate of a threshold value for avalanches of about 7MV/cm (Table I), somewhat dependent on the Xe concentration, and presently limited in accuracy by our knowledge of the needle-mesh gap and the needle tip radius.

One of our tests used pure LAr and we measured needle current over a wide range of voltage. Near the top of the needle the area with surface field values above this threshold, increases with applied voltage. This results in a current approximately proportional to the area above threshold. The result is shown in Fig. 3. The curve shown is a fit to the expected current if there is a threshold avalanche field above which the avalanche is exponential. We model the LAr avalanche multiplication process analogous to the case for gaseous detectors: $I = h \int \exp(\boldsymbol{a}(E/E_{th}-1))d z^2$, where E is the local electric field at the surface, $E_{th}$ is the threshold electric field for avalanches, α is a constant, and h is an overall normalization related to the source intensity and location. The integral is taken over the tip of the needle where E is greater than or equal to $E_{th}$. The ratio of the fields $E/E_{th}$ is related to the voltage on the mesh by: $E/E_{th} = V/(V_{th}(1+p^2(1/z^2-1))^{1/2})$ , where V is the voltage on the mesh and $V_{th}$ is the threshold voltage for avalanches.

Table I. Fields at surface of needle and mesh for the geometry  of  Fig. 2 and 1500V on the mesh.

|  | Field at needle | Field at mesh |
|---|---|---|
| Numeric Calculation ( Maxwell ) | 7.5 MV/cm | 2.55 kV/cm |
| Analytic Calculation | 6.9 MV/cm | 2.57 kV/cm |

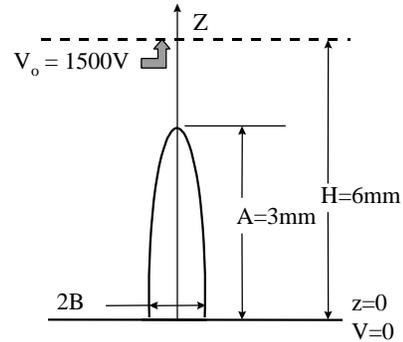

Fig. 2.  The definition of the dimensions of the prolate spheroid

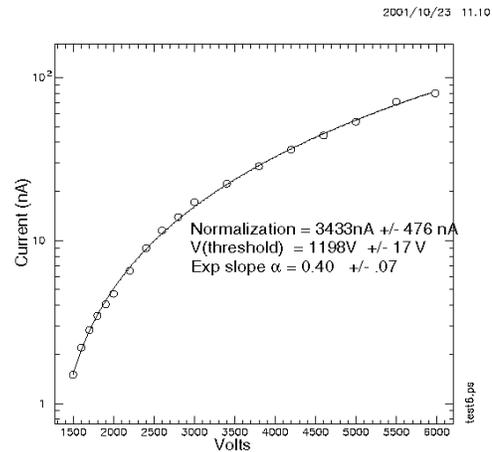

Fig. 3.  Current vs. voltage for a test using pure liqu___    ___rve for a model of area on tip of needle above an avalanche threshold, times an exponential dependence on voltage above threshold. The shape is dominated by the increase in area above a threshold voltage, the increase from avalanche is smaller. The data is for a 4mm high needle in a 6 mm gap, slightly different than the geometry of fig 2.

We fit this integral to the measured values of current versus voltage. The fit depends on three parameters: the arbitrary normalization h, a threshold voltage $V_{th}$ and the exponential slope $\alpha$.

### C.  Avalanche Gain

We have observed avalanche pulse heights between about ~100mV and 3V after amplification, depending upon voltage, pressure, and Xe concentration. Figure 4 is an example of pulses seen on the output of the amplifier chain.

The background noise level is approximately 10-20mV peak-to-peak. Assuming that the 60keV gamma conversions release ~2600 electrons in the liquid argon (23 eV/ion pair), and using the known gain of the amplifier system, the estimated avalanche gain for the smaller pulses in Fig. 4 is ~100, assuming no ion recombination. Since we expect recombination to be a quite significant effect at these fields (20%) [17], the actual avalanche gain is expected to be larger.

### D.  Two Avalanche Modes

As is clearly seen in Fig. 4, there are for some conditions two quite distinct pulse amplitudes, whose heights do not change significantly with source distance. As shown in Fig. 5b, the heights of the larger pulses decrease with He pressure over the range 0 to 1 atm (0-15 psig).

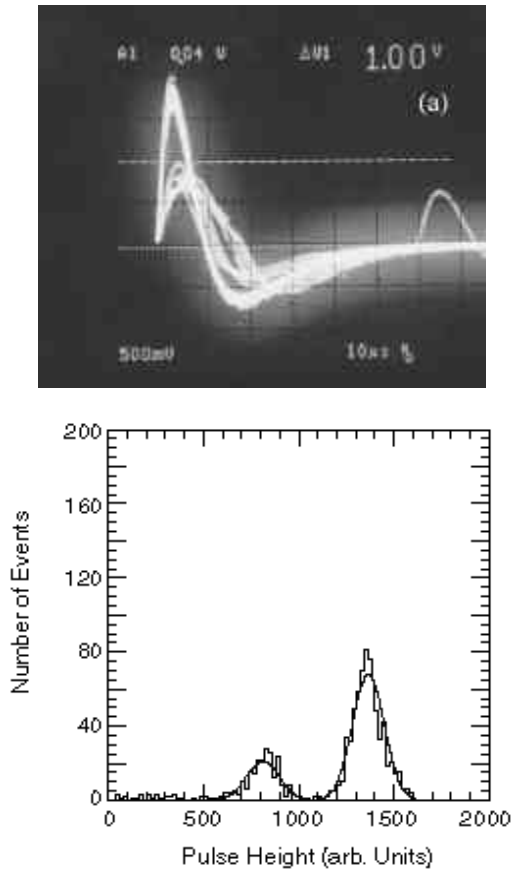

Fig. 4. (a) Photograph of oscilloscope traces of several avalanche events showing two distinct pulse heights. (50ppm Xe, pressure 5psig, HV=1750V, one large division of X axis is 10μs, one large division of Y axis is 500mV); (b) Pulse height distribution for the signals shown in (a).

The smaller pulses, however, are independent of the He pressure. The pressure dependence of the larger pulses suggests that at least some portion of the avalanche mechanism involves the vapor phase (a "bubble" at the needle tip). However, we have no independent verification of this conjecture. Likewise, it is suggestive that the pressure-independent pulses are from avalanches produced entirely in liquid.

### E.  Dependence on Voltage, Pressure and Source Position

Figs. 5a and 5b show the dependence of pulse height on voltage and pressure for each type of pulse shown in Fig. 4:

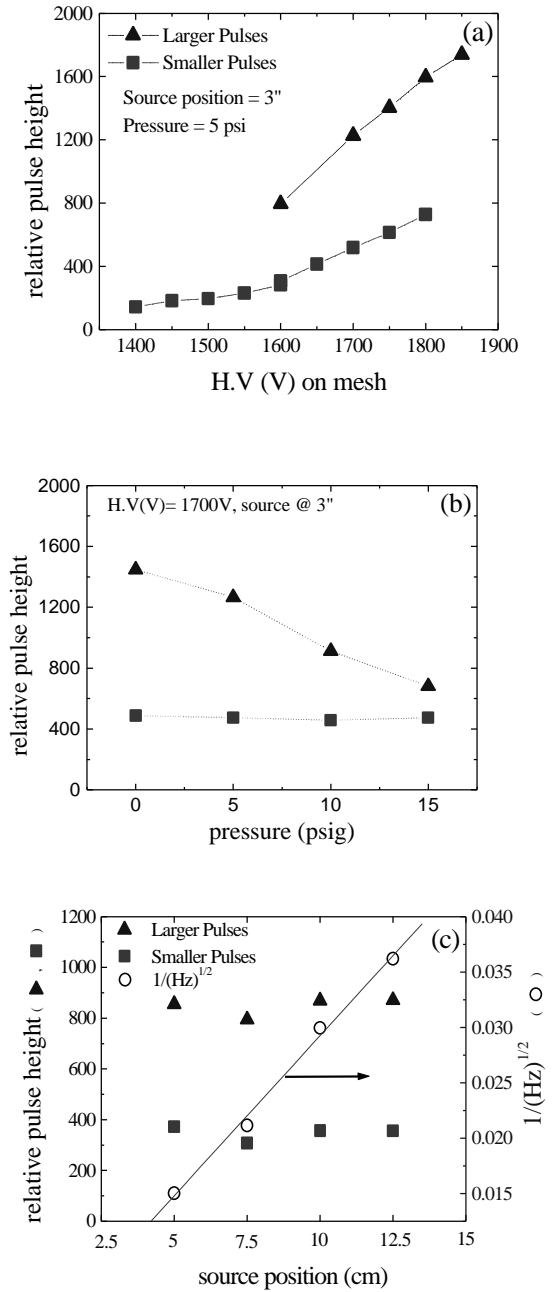

Fig. 5. (a) Pulse height vs. H.V for larger and smaller pulses. (b) Pulse height vs. pressure of He for larger and smaller pulses. (c) Pulse height and $1/(\text{rate})^{1/2}$ vs. source position. The rate is the sum of larger and smaller pulses (HV =1600 V, pressure = 5psig, Xe/Ar concentration = 100ppm)

the pressure-sensitive pulses, and the smaller, pressure-independent pulses. Fig. 5c shows that the pulse height of both types of pulses are independent of the source position while the total rate of both pulses agrees well with a $1/(distance)^2$ dependence. Generally, both types of pulses are quite uniform in shape, and have rather narrow pulse height distributions, as shown in Fig. 4. The large pulses are well below the saturation voltage of the pre-amplifier or post-amplifier.

### F. Gain vs. Xenon concentration

We have explored the effects of Xe for eight Xe/Ar concentrations in the range of 25 to 50,000ppm. In pure Argon, the behavior is unstable, with the pulse amplitude and rate changing over tens of seconds, or disappearing altogether, only to re-appear a few seconds later. Similar behavior is reported in ref [3]. Avalanche behavior stabilizes with the addition of even small amounts of Xe. Fig. 6 shows the pulse height (PH) at 1500V vs. Xe concentration. An interesting behavior is that the PH becomes smaller between 25ppm and 500ppm, then becomes larger again, still increasing at the greatest concentration, 50,000ppm. An increase in ionization density for LAr with Xe concentrations above 500ppm was reported in [17], but it was always less than 15%, and not in avalanche mode.

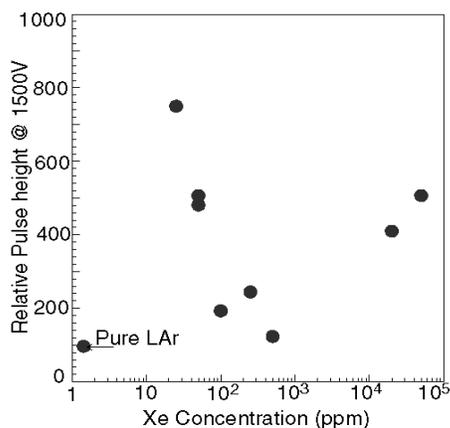

Fig. 6. Relative pulse height of smaller, pressure independent pulses as a function of xenon concentration at field strengths ~9x10⁶V/cm.

The pressure-dependent (larger) pulses were observed in the small xenon concentration region (50ppm to 500ppm), but no longer appeared at large xenon concentrations (1000ppm to 50,000ppm). In several tests we observed a difference in performance as a function of time: typically, the PH decreased somewhat as the rate increased (or vice versa) over a period of several hours. This behavior, which occurred primarily in the low concentration region, is not yet understood, but may come from inhomogeneites in the Xe concentration.

### G. Time Decay Spectra vs. Xenon Concentration

Processes which are known to occur in liquid Xe/Ar mixtures include the transfer of energy from ionized Ar to Xe because of the higher ionization potential for LAr (14.2eV

[18]) than for LXe (10.6eV for Xe dissolved in LAr [17]). Other processes are the formation, and the subsequent de-excitation of $Ar*_2$ or $Xe*_2$ excimers. These processes lead to emission of scintillation light in the far UV region, peaking at 128 nm from LAr, and 175 nm from LXe.

Time spectra were taken by using the needle avalanche as the start pulse on a time-to-pulse-height converter, and the detection of a UV photon in the PMT with a CsTe photocathode (or gaseous detector, see below) as the stop pulse. The decay time is determined by the de-excitation of the Xe/Ar system. An example of such a spectrum is shown in Fig. 7. Separate tests indicate the system is capable of measuring decay constants as short as ~15nsec.

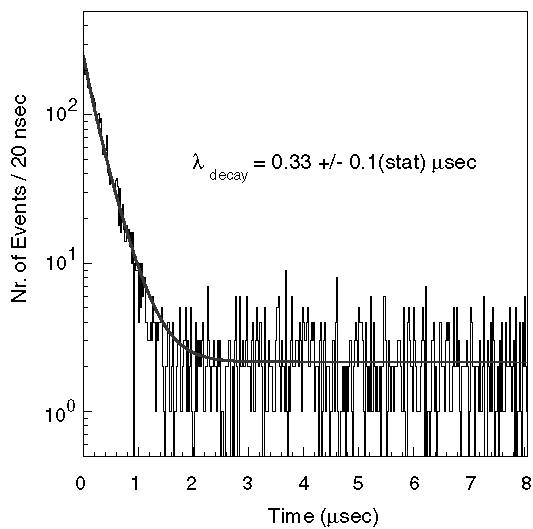

Fig. 7. Example of a time distribution for coincidences between the needle and the PMT. The data are fit (smooth curve) as described in the text, with a decay constant as indicated. The Xe/Ar concentration is 2000ppm.

For the data shown here, the average number of photons ($\lambda$) observed per event ranged between $0 < \lambda < 1.5$, depending on the test and conditions, To extract the lifetime, $\tau$, we fit the decay time distribution with the function $F(t) = B \exp(-\lambda(1 - e^{-t/t}))e^{-t/t} + k$, where B is a constant, $\lambda$ is the measured, average number of photons/event, and the constant k describes the flat background. The first exponential represents the probability that a photon does not appear before time t, and the last exponential is the probability density for a photon appearing at time t.

Shown in Fig. 8 is comparison of time decay constants measured in our experiment with that of Kubota, et al., [19] and Conti, et al., [20]. It is evident that we have decay times more consistent with those measured by Conti, et al. This result appears to be consistent with the fact that both that experiment and ours are sensitive only to the 175nm wavelength of light from Xe, but insensitive to Ar light. We have no explanation for the anomalously low time constant for the one point at 50ppm Xe, compared to the rest of our data. Understanding this point will require further investigation.

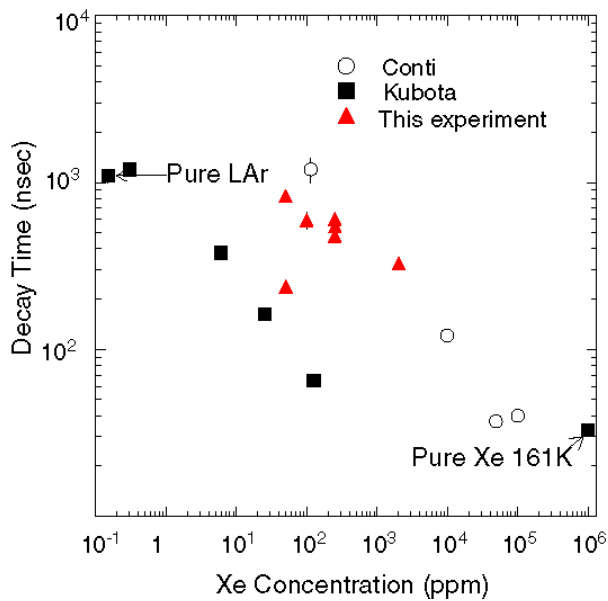

Fig. 8. A comparison of our measured decay time constants with those of Kubota, et al.[19], and Conti, et al. [20].

### H.    Other Measurements

In the course of our investigations we have made other measurements not discussed in detail here. We were successful in producing avalanches in LAr containing ~4ppm TMS (tetramethylsilane), but using electron dispersive x-ray fluorescence (EDXF) surface analysis we found silicon deposits near the tip of the needle after the test. Similar attempts with (30%) neon in LAr have been unsuccessful in producing reliable avalanche behavior.

We could easily detect photons from the decay of Xe*$_2$ (Ar*$_2$) excimers using a gaseous detector with a CsI photocathode, avalanching the photoelectrons on a 50μm tungsten wire in P-10 (CF$_4$) gas. The time resolution was not as good as that of the PMT because of the variation of the drift time of the photoelectrons in the gas.

Using a recently acquired 10mCi $^{57}$Co source (maximum photon energy 135 keV) illuminating the needle through the 0.25mm thick Al window in the test vessel, we obtained pulse heights identical to those obtained with the $^{241}$Am source (maximum photon energy 60 keV), in LAr with ~250ppm Xe when both are measured at the same voltage. The presence of the $^{57}$Co source is distinguished only by the factor of ~10 higher rate of signals it induces on the needle, and by the much larger photon energy than for $^{241}$Am. This is our first and only test to date with the $^{57}$Co source, and this test implies that the pulses are saturated relative to the ionization initiating the avalanche, since the pulse heights do not depend on the incident photon energy. Our conjecture is that we may be seeing limited avalanche, or 'streamer', behavior, which are saturated avalanche modes observed in gas avalanche chambers. However, the current versus HV behavior (Fig. 3) is inconsistent with Geiger mode.

We have also tested 0.25 μm and 0.40μm wide chromium anodes interdigitated at a pitch of 400μm with 5μm wide cathodes deposited on a quartz substrate. We saw small pulses (50 to 100mV) that gradually disappeared over a few minutes. We suspect the quartz substrate is charging up, and canceling the electric field. Avalanche amplification of drifting electrons has been seen on microstrip detectors in pure liquid Xe [21] with an amplification factor of ~10.

### IV.    Conclusions

We have found that small concentrations of Xe mixed with LAr stabilize the electron avalanche process, which is erratic in pure liquid argon. Tests have been done over a wide range of Xe/Ar concentrations to characterize the avalanche behavior. The signal/noise ratio was much larger than 100:1, and avalanche gains are estimated to be ~100. We find that there are two modes, one with pressure-dependent signal pulses, and another with smaller, pressure-independent pulses. Both types of pulses have rather narrow distributions and may be saturated, i. e. not proportional to primary ionization. The pulse height increases approximately linearly with applied voltage. A $^{57}$Co source gave the same pulse amplitude and spectra as $^{241}$Am, even though the photon energy is much larger for the $^{57}$Co, a strong indication of saturation. Time spectra of photon emission are in agreement with measurements done by others, and are sufficiently fast for a trigger signal for many applications. Experiments requiring large volumes of Ar would not be limited by the cost of the Xe.

### Acknowledgments

We would like to thank S. Derenzo for several informative discussions, and J. Wise for clarifying Xe solubility in Ar. Also, we thank E. Saiz for help with the electron microscope facility, and professors K. Abe and H. Yoshimoto for supplying samples of TMS and TMG.